\documentclass[pra,aps,twocolumn,superscriptaddress,amsfonts,citeautoscript,nofootinbib,a4paper]{revtex4}

\usepackage{amsmath}  \usepackage{amssymb}  \usepackage{amsfonts}  \usepackage{bm}  \usepackage{bbm}   \usepackage{braket}  \usepackage{color}  \usepackage{comment}  \usepackage{dcolumn}  \usepackage{enumerate}  \usepackage{epsfig}  \usepackage{gensymb}  \usepackage{graphicx}  \usepackage{indentfirst}  \usepackage{lmodern}  \usepackage{mathrsfs}  \usepackage{mathtools}  \usepackage{psfrag}  \usepackage{pst-all}  \usepackage{soul}  \usepackage{units}  \usepackage{xcolor}
\usepackage[colorlinks,linkcolor=blue,citecolor=blue,urlcolor=blue,hyperindex,driverfallback=dvipdfm]{hyperref}  \usepackage[T1]{fontenc} %---APS---ACS---SI---arxiv

\def\ii{{\rm i}}  \def\ee{{\rm e}}
\def\me{m_{\rm e}}  
  \def\Imm{{\rm Im}}
          \def\eb{{\bf e}}                              \def\Qb{{\bf Q}}  \def\qb{{\bf q}}  \def\Rb{{\bf R}}  \def\rb{{\bf r}}  \def\Sb{{\bf S}}    \def\vb{{\bf v}} %--- bold vectors
    \def\zz{\hat{\bf z}}            
\def\epsilonb{\epsilon_b}

\begin{document} %---APS---SI---arxiv
\def\bibsection{\section*{\refname}} %---SI---arxiv

\title{Entangling free electrons and optical excitations}

\author{Andrea Kone\v{c}n\'{a}}
\affiliation{ICFO-Institut de Ciencies Fotoniques, The Barcelona Institute of Science and Technology, 08860 Castelldefels (Barcelona), Spain}
\affiliation{Central European Institute of Technology, Brno University of Technology, Brno 61200, Czech Republic}
\author{Fadil Iyikanat}
\affiliation{ICFO-Institut de Ciencies Fotoniques, The Barcelona Institute of Science and Technology, 08860 Castelldefels (Barcelona), Spain}
\author{F.~Javier~Garc\'{\i}a~de~Abajo}
\email{javier.garciadeabajo@nanophotonics.es}
\affiliation{ICFO-Institut de Ciencies Fotoniques, The Barcelona Institute of Science and Technology, 08860 Castelldefels (Barcelona), Spain}
\affiliation{ICREA-Instituci\'o Catalana de Recerca i Estudis Avan\c{c}ats, Passeig Llu\'{\i}s Companys 23, 08010 Barcelona, Spain}

\begin{abstract}
The inelastic interaction between flying particles and optical nanocavities gives rise to entangled states in which some excitations of the latter are paired with changes in the energy or momentum of the former. In particular, entanglement of free electrons and nanocavity modes opens appealing opportunities associated with the strong interaction capabilities of the electrons. However, the degree of entanglement that is currently achievable by electron interaction with optical cavities is severely limited by the lack of external selectivity over the resulting state mixtures. Here, we propose a scheme to generate pure entanglement between designated optical excitations in a cavity and separable free-electron states. Specifically, we shape the electron wave-function profile to dramatically reduce the number of accessible cavity modes and simultaneously associate them with targeted electron scattering directions. We exemplify this concept through a theoretical description of free-electron entanglement with degenerate and nondegenerate plasmon modes in silver nanoparticles as well as atomic vibrations in an inorganic molecule. The generated entanglement can be further propagated through its electron component to extend quantum interactions beyond currently explored protocols.
\end{abstract}

\maketitle
\date{\today}

\section{Introduction}

Although entangled states in the context of quantum optics are generally relying on photons \cite{HHH09,TCT10}, the exploration of entanglement with other types of information carriers could open a wealth of possibilities to discover new phenomena and materialize disruptive protocols for quantum metrology and microscopy \cite{K19,paper339,RML20}. In particular, free electrons are advantageous candidates because they can undergo substantial inelastic scattering by nanostructures \cite{paper149}, which is an attribute enabling electron energy-loss spectroscopy (EELS) performed in electron microscopes to reveal the presence, strength, and spatial distribution of optical excitations down to the atomic scale \cite{E96,E03,EB05,B06,KLD14,KDH19,paper371}. Actually, low-loss EELS has been extensively used to study atomic vibrations in low-dimensional materials \cite{HNY18,HRK20,YLG21} and molecules \cite{RAM16,HC18,JHH18,HHP19}, collective excitations such as plasmons \cite{BKW07,paper085,RB13,FWY14,paper369} and phonon polaritons \cite{KLD14,LTH17,GKC17,paper361}, and photon confinement in optical cavities \cite{KLS20,WDS20,paper383}.

In momentum-resolved EELS, each excitation event produced by a traversing electron is individually identified through an electron measurement as a function of the deflection angle and energy loss, and therefore, this configuration already generates entanglement between electron states with different energy/momentum and excitations in the sampled structure. Consequently, the post-interaction electron-sample state has the form
\begin{align}
|\Psi_f\rangle=\sum_n\int d^2\Qb_f\;\alpha_{\Qb_f n}^f\,|\Qb_f\rangle\otimes|n\rangle,
\label{psif}
\end{align}
where $n$ and $\Qb_f$ run over final sample and electron-wave-vector states, respectively, and $\alpha_{\Qb_f n}^f$ are complex scattering amplitudes \cite{paper371}. But unfortunately, the resulting electron-sample mixture of states is generally too complex to be of practical interest for quantum technologies. Nevertheless, this approach holds elements of novelty with respect to traditional quantum optics methods because one of the entangled particles (the free electron) can be highly energetic, and therefore capable of undergoing subsequent strong collisions with other objects.

Free-electron waves can be manipulated with great precision thanks to an impressive series of advances occurred in electron microscopy over the last decades. Currently, electron beams (ebeams) can be collimated and focused with sub-{\aa}ngstrom spatial precision \cite{BDK02}, monochromatized within a few meV \cite{KLD14,LTH17}, and temporally compressed down to femtosecond \cite{BFZ09,FES15,PLQ15} and even attosecond \cite{PRY17,KSH18,MB18_2} time scales. In addition to traditional electron-optics lenses \cite{CBK13}, control over the transverse electron wave function can be exerted by means of beam splitters \cite{MD1956,GBL17}, engineered gratings \cite{JTM21,JTG21}, chiral transmission masks \cite{UT10,VTS10,MAA11}, magnetic monopole fields \cite{BVV14}, electrically programmable phase plates \cite{VBM18}, and active optical-phase imprinting \cite{paper311,paper332,SAC19,FYS20,paper351,paper368}. A vibrant community is swiftly gathering around these methods, which are the basis for elastic \cite{HS1972,H08_2} and inelastic \cite{LF00,PLV06,VBS08} holography, and further enable the synthesis of vortex ebeams \cite{UT10,VTS10,MAA11,BIG17,paper332}, the study of magnetic \cite{VTS10,RI16} and optical dichroism \cite{paper243,ZRF19,GRZ21}, and the excitation of localized optical modes of selected symmetry \cite{GBL17}.

The manipulation of the longitudinal electron wave-function component is also possible in ultrafast electron microscopes \cite{ARM20}, where femtosecond electron pulses are produced from photocathodes illuminated by pulsed lasers, and the subsequent synchronized light-electron interaction allows one to inspect the specimen with femtosecond time resolution. This is the so-called photon-induced near-field electron microscopy \cite{BFZ09,paper151,FES15,PLQ15,KLS20,WDS20,HRF22} (PINEM), which, combined with free propagation, leads to attosecond electron compression \cite{BZ07,PRY17,MB18_2,MB20} and endows the free electrons with the ability to transfer quantum coherence between different systems \cite{paper374,paper373}. The field is thus ripe for the exploitation of free electrons as additional elements in the quantum technology Lego, but as impressive as these advances may seem, they have not yet been leveraged to generate pure entanglement between light and free electrons.

% Figure 1 ------------------------------------------------
\begin{figure}
\centering{\includegraphics[width=0.48\textwidth]{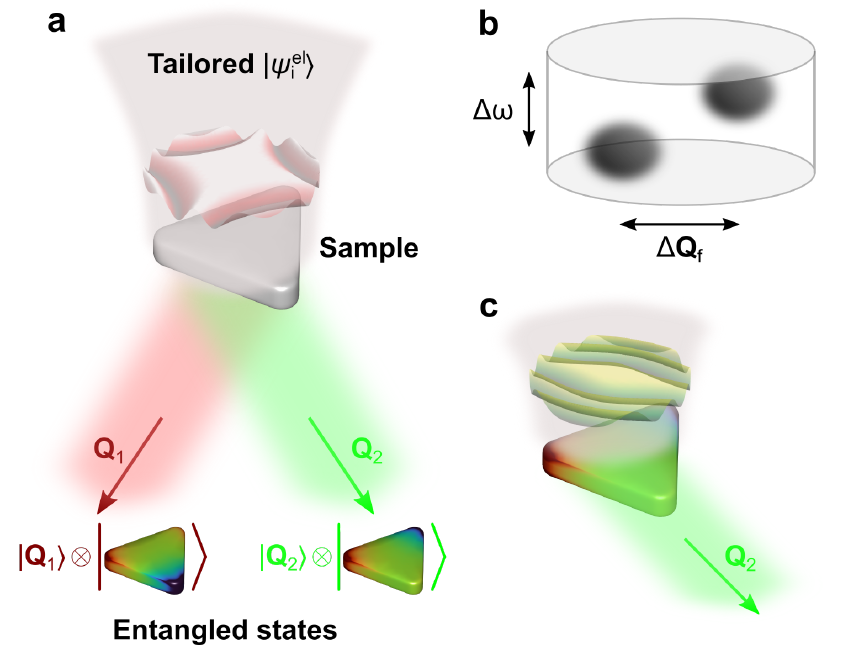}}
\caption{{\bf Proposed scheme for the generation of entangled electron-cavity states.} (a) A preshaped electron interacts with a nanostructure (a triangular plasmonic cavity) supporting well-defined optical or vibrational modes. The incident electron wave function $\ket{\psi_i^{\rm el}}$ is tailored such that we obtain entangled states after interaction, correlating different sample excitations (colored triangles) with separated electron scattering directions (final electron state having components of transverse wave vectors $\Qb_1$ and $\Qb_2$). A maximally entangled electron-sample state is thus produced, as the sample is in a superposition of excited states correlated with different electron scattering directions. (b) Electrons are emerging along separate spots within a finite region of size $\Delta\hbar\omega\times\Delta\hbar\Qb_f$ in the configuration space of energy-loss and transverse-momentum transfers. (c) Momentum-filtering at the electron detector allows us to project on the desired sample mode and eventually explore its dynamics through subsequent interrogation, for example by exposure to a synchronized light pulse.}
\label{Fig1}
\end{figure}

Here, we demonstrate through rigorous quantum theory that pure entanglement between electrons and confined optical modes can be generated by suitably patterning the transverse incident electron wave function. As schematically illustrated in Fig.~\ref{Fig1}a, the electron undergoes a change in the direction of propagation after being inelastically scattered by the sample, and we prepare the incident electron phase profile in such a way that only a few sample excitations are accessible (two in the figure), leading to separable transmission directions (transverse wave vectors $\Qb_1$ and $\Qb_2$). The two possible excitations created by the electron and their different associated scattering directions form a maximally entangled state. In essence, we specify a finite volume in the configuration space of transmitted electrons defined by an energy-loss window $\Delta\hbar\omega$ and a transverse momentum area $\Delta\hbar\Qb_f$ in which the final state only populates two well-defined spots (Fig.~\ref{Fig1}b). As we demonstrate below, this approach can be also used to create heralded single sample excitations (Fig.~\ref{Fig1}c). In addition, manipulation of the electron component in electron-sample entangled states through, for example, electron interference could be used to process quantum information and imprint it on other (eventually macroscopic) objects via subsequent interactions.

% =========================================================
% --- results ... -----------------------------------------
% =========================================================
\section{Results and Discussion}

% --- theory ----------------------------------------------
\subsection{Free-electron interaction with confined optical modes}

We intend to synthesize an electron-sample state as described by Eq.~\eqref{psif}, with the free-electron component piled up at separate regions in momentum-energy space (Fig.~\ref{Fig1}b) and a different sample excitation associated with each of those regions. The starting point is the initial combined state
\begin{align}
|\Psi_i\rangle=|\psi_i^{\rm el}\rangle\otimes|0\rangle,
\nonumber
\end{align}
where the sample is in its ground state $\ket{0}$ and the incident electron wave function, whose spatial dependence is given by
\begin{align}
\psi_i^{\rm el}(\Rb)=\int d^2\Qb_i\;\alpha_{\Qb_i}^i\big(\ee^{\ii\Qb_i\cdot\Rb}/2\pi\big), \label{psiiel}
\end{align}
is prepared as a combination of momentum states with coefficients $\alpha_{\Qb_i}^i$ determined through the use of customized transmission masks \cite{UT10,VTS10,MAA11} or phase imprinting based on electrostatic \cite{VBM18} and optical \cite{paper351,paper368} fields. We consider incident monochromatic electrons, so that the dependence of the electron wave function on 2D transverse coordinates $\Rb$ and its decomposition in 2D wave vectors $\Qb_i$ is everything we need to describe the electron in the interaction region without loss of generality.

Electron-sample interaction operates a linear transformation relating the final coefficients $\alpha_{\Qb_f n}^f$ in Eq.~\eqref{psif} to $\alpha_{\Qb_i}^i$ in Eq.~\eqref{psiiel}. More precisely,
\begin{align}
\alpha_{\Qb_f n}^f=\int d^2\Qb_i\;M_{\Qb_f-\Qb_i,n}\;\alpha_{\Qb_i}^i,
\label{aMa}
\end{align}
where $M_{\Qb_f-\Qb_i,n}$ only depends on the momentum transfer $\hbar(\Qb_i-\Qb_f)$ for each excited state $n$ (see Appendix).

A connection can be readily established with EELS experiments, in which electron counts are recorded as a function of the energy loss $\hbar\omega$, thus yielding a frequency- and momentum-resolved loss probability $\Gamma_\mathrm{EELS}(\Qb_f,\omega)=\sum_n\big|\alpha_{\Qb_f n}^f\big|^2\;\delta(\omega-\omega_n)$, where $\hbar\omega_n$ is the excitation energy of sample mode $n$. Within first-order perturbation theory, and further adopting the electrostatic and nonrecoil approximations, the angle-resolved EELS probability can be expressed in terms of mode-dependent dimensionless spectral functions $g_n(\omega)$ as
\begin{align}
\Gamma_\mathrm{EELS}(\Qb_f,\omega)&=\frac{e^2}{4\pi^3\hbar v^2}\sum_n g_n(\omega) \label{Eq:dGamdQf1}\\ 
&\times\left\lvert \int d^2\Rb\,\psi_i^{\rm el}(\Rb)\ee^{-\ii\Qb_f\cdot \Rb}\,w_n(\Rb,\omega)\right\rvert^2,
\nonumber
\end{align}
where $v$ is the electron velocity and
\begin{align}
w_n(\Rb,\omega)\propto\int d^2\Qb\;\ee^{-\ii\Qb\cdot \Rb}\;M_{\Qb n} \label{Mvsw}
\end{align}
gives the spatial profile of mode $n$ (see details in the Appendix, including expressions of the quantities $g_n(\omega)$ and $w_n(\Rb,\omega)$ associated with plasmons and atomic vibrations).

Here, we are interested in determining the incident electron wave-function profile (i.e., the momentum-dependent coefficients $\alpha_{\Qb_i}^i$) such that different sample modes $n$ are associated with final wave-function coefficients $\alpha_{\Qb_f n}^f$ within well separated regions in momentum space (see Fig.~\ref{Fig1}b). To demonstrate the feasibility of this concept in the synthesis of electron-sample entanglement, we invert Eq.~\eqref{aMa} with a predetermined choice of $\alpha_{\Qb_f n}^f$, which we set to designated values for each sample excitation $n$ within a targeted finite-size region in $\Qb_f$ space (see details in the Appendix). This simple procedure is sufficient for the proof-of-principle demonstration that we pursue in this work. However, more elaborate schemes for incident electron wave-function optimization could rely on iterative methods or neural-network training \cite{SOJ21}.

% Figure 2 ------------------------------------------------
\begin{figure}
\centering{\includegraphics[width=0.4\textwidth]{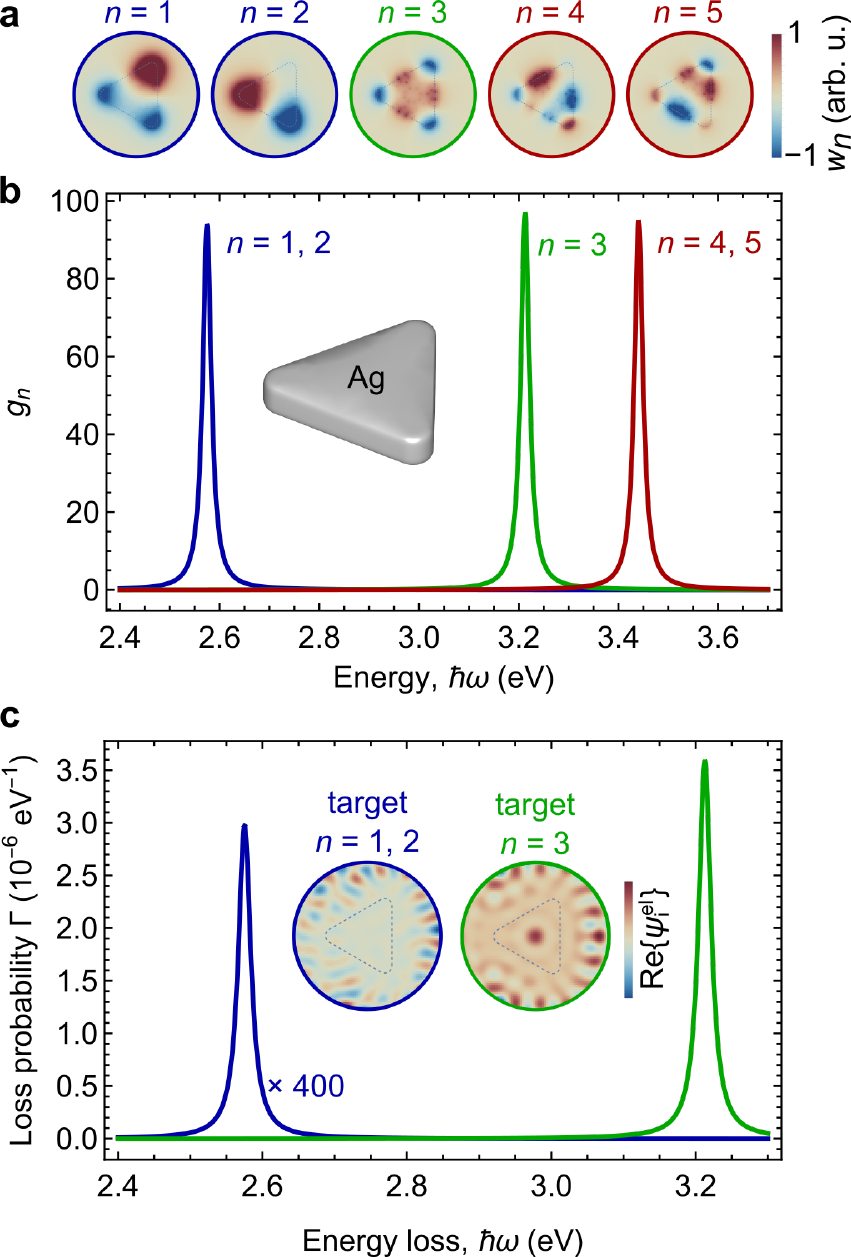}}
\caption{{\bf Selective excitation of plasmon modes in a silver nanotriangle.} (a,b) Spatial profiles (a) and spectral functions (b) associated with plasmons in a silver nanotriangle (2\,nm thickness, 10\,nm side length). We find two sets of degenerate modes (left and right peaks) and one nondegenerate mode (see color-matched labels with the index $n$). (c) Electron energy-loss spectra for two optimized incident electron wave-function profiles $\psi_i^{\rm el}(\Rb)$, the real part of which is represented as a function of transverse coordinates $\Rb$ in the insets, framed in color-matched circumferences. The optimization is carried out for 100\,keV electrons, an electron detector consisting of 49 pixels, an incident convergence half-angle $\varphi_i=1.5\,$mrad, and a collection half-angle $\varphi_f=0.75\,$mrad. The nanotriangle contour is indicated by thin dashed curves in (a) and (c).}
\label{Fig2}
\end{figure}

% Figure 3 ------------------------------------------------
\begin{figure*}
\centering{\includegraphics[width=0.75\textwidth]{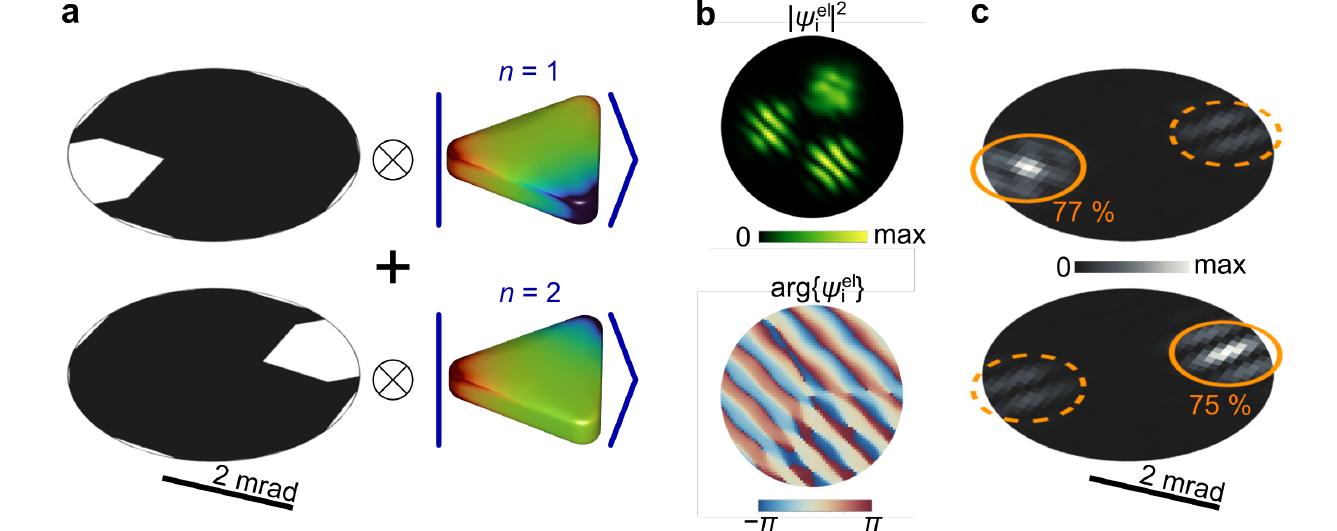}}
\caption{{\bf Creation of electron-sample states with a high-degree of entanglement.} (a) Pursued electron-sample entangled state, consisting of the superposition of selected electron-momentum states within the white pixels in $\Qb_f$ space (left) and correlated degenerate dipolar plasmons in the silver nanotriangle sample considered in Fig.~\ref{Fig2} (right). (b) Spatial profile of the optimized incident wave function $\psi_i^{\rm el}(\Rb)$ required to produce the final state in (a). (c) Resulting probability distributions $\big|\langle \Qb_f,n|\Psi_f\rangle\big|^2$ with $n=1$ (top) and $n=2$ (bottom) in $\Qb_f$ space, where the colored circles are composed of the indicated fractions of the targeted excitation $n$. The optimization is carried out for 100\,keV electrons, 13 detector pixels, $\varphi_i=4\,$mrad, and $\varphi_f=2\,$mrad.}
\label{Fig3}
\end{figure*}

% Figure 4 ------------------------------------------------
\begin{figure}
\centering{\includegraphics[width=0.45\textwidth]{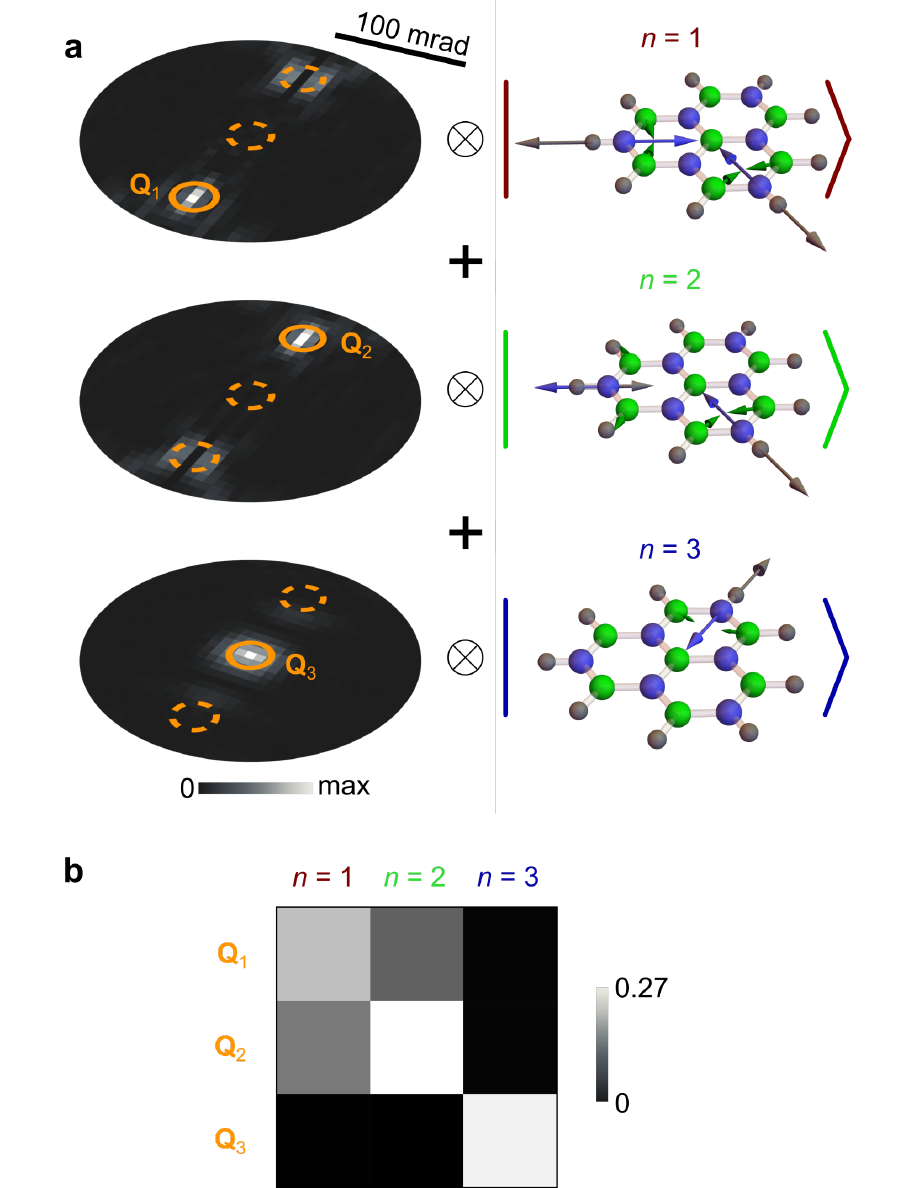}}
\caption{{\bf Entanglement of free electrons and atomic vibrations.} (a) Representation of the obtained electron-sample entangled state. We plot the momentum distributions of scattered electrons (left) corresponding to the excitation of the 440\,meV triply-degenerate vibrational modes of a hBN molecule (right). (b) Probability matrix showing the fractional contribution associated with the excitation of each of the three vibrational modes $n=1-3$ to the energy-filtered electron signal contained within the three selected circular areas around final transverse wave vectors $\Qb_1$, $\Qb_2$, and $\Qb_3$ in (a). The sum of the 9 matrix elements is normalized to 1. We consider 60\,keV electrons, 29 detector pixels, and $\varphi_i=\varphi_f=100\,$mrad.}
\label{Fig4}
\end{figure}

% --- plasmon-mode selection ------------------------------
\subsection{Selected excitation of individual plamons}

As a preliminary step before addressing electron-sample entanglement, we tackle the problem of selectively exciting a single plasmon in a metallic nanoparticle. Although this can be achieved through post-selection of a small range of scattered electron wave vectors \cite{GBL17}, we formulate a solution in which the plasmon-exciting electrons emerge within a relatively large region in momentum space, and this solution is generalized below to create entanglement. We consider a silver triangle that sustains five plasmon modes in the 2.4-3.7\,eV spectral region \cite{SDH14_2}: two sets of doubly-degenerate dipolar (blue curve and circles, $n=1,2$) and quadrupolar (red, $n=4,5$) plasmons, and one nondegenerate hexapolar mode (green, $n=3$), as revealed by the spatial and spectral functions plotted in Fig.~\ref{Fig2}a,b (see details of the calculation in the Appendix). We then optimize the incident electron wave function over a $\Qb_i$ region discretized with 1257 pixels and defined by a convergence half-angle $\varphi_i=1.5\,$mrad, such that either $n=1,2$ or $n=3$ are the only modes excited when the scattered electrons are collected over a $\Qb_f$ region spanning a half-angle $\varphi_f=0.75\,$mrad (discretized with 49 pixels) and energy-filtered between 2.4 and 3.3\,eV.

The resulting real-space profiles of $\psi_i^{\rm el}(\Rb)$ are shown in the insets of Fig.~\ref{Fig2}c (circular color plots), along with the color-matched EELS probability curves obtained from Eq.~\eqref{Eq:dGamdQf1} by collecting only electrons that emerge within the indicated $\Qb_f$ and energy region. Incidentally, all modes are excited by the incident electron because they have overlapping spatial distributions (Fig.~\ref{Fig2}a) and the EELS probability integrated over all possible $\Qb_f$'s is rigorously given by the incoherent average over incident electron positions $\Rb$, weighted by the electron probability \cite{RH1977,paper149} $\big|\psi_i^{\rm el}(\Rb)\big|^2$ (see Appendix). But remarkably, our simple optimization procedure is capable of placing the weight of the excitation of either $n=1,2$ or $n=3$ modes preferably inside the $\Qb_f$ region defined by a collection half-angle $\varphi_f=0.75\,$mrad, while electrons producing either $n=3$ or $n=1,2$, respectively, are left outside that region.

% --- entanglement with plasmons --------------------------
\subsection{Generation of electron-plasmon entangled states}

We now apply the principle of $\psi_i^{\rm el}$ shaping to demonstrate the generation of electron-sample entanglement for the same triangular sample as considered above. Specifically, we focus on the lowest-energy degenerate plasmons $n=1,2$ and aim at correlating these excitations with final electron momentum states along separate $\Qb_f$ directions (Fig.~\ref{Fig3}a). Following the same procedure as above, we find the optimized electron wave function shown in Fig.~\ref{Fig3}b, from which we obtain the actual scattered electron distribution plotted in Fig.~\ref{Fig3}c in $\Qb_f$ space for components corresponding to the excitation of $n=1$ (top) and $n=2$ (bottom) modes. When examining the $\Qb_f$ region enclosed by the two colored circles in Fig.~\ref{Fig3}c, we find that 77\% of the electron signal inside the left one is associated with the excitation of the $n=1$ plasmon, whereas the right circle is made of 75\% excitation of $n=2$, thus revealing a high degree of entanglement between the excited plasmons and the selected electron scattering directions. We note that the symmetry of the selected degenerate plasmons plays a similar role as photon polarization in light-based entanglement schemes \cite{HHH09}.

% --- vibrational states ----------------------------------
\subsection{Electron entanglement with atomic-vibrational states}

The electron-sample entanglement scheme under consideration can be applied to sample excitations of different nature. We illustrate this versatility by considering atomic vibrations in a hexagonal boron nitride (hBN) molecule (Fig.~\ref{Fig4}), which we simulate from first principles \cite{paper376} (see Appendix) assuming passivation of the edges with hydrogen atoms. This structure supports a number of excitations up to energies $\sim450\,$meV, including a set of triply-degenerate N-H bond-stretching modes at $440\,$meV, on which we focus our analysis. We again optimize the incident electron wave function to achieve entanglement between final electron states and vibrational modes of the molecule. Because of the strong spatial confinement of vibrational modes, the angular ranges that need to be considered for the incident and scattered electron wave functions are now considerably larger than for plasmons (cf. angle scales in Figs.~\ref{Fig3} and \ref{Fig4}). The achieved electron-sample state, illustrated in Fig.~\ref{Fig4}a, exhibits a high degree of entanglement when selecting electrons scattered along the colored circles in $\Qb_f$ space, also revealed through the partial probabilities contributed by each of the three vibrational modes to each of the regions enclosed by those circles (see table in Fig.~\ref{Fig4}b).

% =========================================================
% --- conclusions -----------------------------------------
% =========================================================
\section{Concluding Remarks}%---APS---OSA---arxiv
%\section*{CONCLUSIONS} ... text ... %---ACS

By entangling the transverse momenta of free electrons with localized optical excitations in a nanostructure, we could selectively measure one of the corresponding outgoing electron directions, thus providing a way to herald the creation of single designated excitations in the sample. This should allow us to follow the dynamics of the later and gain insight into the state-dependent decay pathways, for example by subsequently probing the evolution of the specimen through scattering of laser pulses that are synchronized with the electron in an electron-pump/photon-probe approach. An additional possibility is offered by correlating the angle-resolved electron signal with traces originating in the decay of the sample excited states (e.g., an electrical signal produced by coupling to electron-hole pairs in a proximal semiconductor or also the polarization- and angle-resolved cathodoluminescence emission associated with radiative decay). The present scheme could also be extended to incorporate gain processes similar to those in PINEM upon illumination of the sample with symmetry-matched optical pulses that can simultaneously excite a subset of its supported excitations. Finally, besides the investigated examples of plasmons in nanoparticles and atomic vibrations in molecules, free electrons could also be entangled with optical modes in dielectric cavities \cite{paper383} and photons guided along optical waveguides \cite{paper180}, which together configure a vast range of possibilities for leveraging the quantum nature of free electrons in the design of improved microscopy and metrology schemes.

\appendix
\section*{APPENDIX}

\section{Transfer matrix for inelastic electron-sample scattering}

The time-dependent electron-sample system can be generally described by a wave function of the form $\ket{\psi(t)}=\sum_{n}\int d^3\qb\;\alpha_{\qb n}(t)\ee^{-\ii(\epsilon_\qb+\omega_n)t}\ket{\qb}\otimes\ket{n}$, where $\ket{\qb}$ and $\ket{n}$ are 
electron and sample eigenstates of the noninteracting Hamiltonian with energies $\hbar\varepsilon_q$ and $\hbar\omega_n$, respectively. In particular, electron states are labeled by the three-dimensional momentum $\hbar\qb$ and satisfy the orthonormality relation $\langle\qb|\qb'\rangle=\delta(\qb-\qb')$. The expansion coefficients $\alpha_{\qb n}(t)$ are determined by solving the Schr\"odinger equation with an electron-sample interaction Hamiltonian $\hat{\mathcal{H}_1}$, which is generally weak for the energetic probes that are typically employed in electron microscopes, so we can work within first-order perturbation theory. Then, taken the sample to be initially prepared in its ground state $n=0$, the post-interaction wave function has coefficients $\alpha_{\qb n}(\infty)=(-2\pi\ii/\hbar)\int d^3\qb'\;\delta(\epsilon_\qb-\epsilon_{\qb'}+\omega_n)\bra{n}\bra{\qb}\hat{\mathcal{H}_1}\ket{\qb'}\ket{0}\alpha_{\qb'0}(-\infty)$, where we set $\omega_0=0$ without loss of generality. We further adopt the nonrecoil approximation \cite{paper371} $\epsilon_\qb-\epsilon_{\qb'}\approx(\qb-\qb')\cdot\vb$ under the assumption that the transverse electron energy is negligible compared with the longitudinal energy along the ebeam direction defined by the average electron velocity $\vb$. This condition is commonly satisfied in electron microscopes. In this approximation, the energy $\hbar\omega_n$ transferred from the electron to the sample is fully absorbed by a change in the longitudinal electron wave vector given by $-\omega_n/v$, so for monochromatic incident electrons, the initial and final longitudinal components of the electron wave function play a trivial role and can be disregarded in the description of the present problem. Consequently, we can expand the final wave function as shown in Eq.~\eqref{psif}, with coefficients $\alpha_{\Qb_f n}^f\equiv\alpha_{\qb n}(\infty)$ that only depend on the transverse electron wave vector $\Qb_f$ for each sample excitation $n$ and are determined from the incident electron wave-function coefficients $\alpha_{\Qb_i}^i\equiv\alpha_{\qb0}(-\infty)$ through the linear relation
\begin{align}
\alpha_{\Qb_f n}^f=\int d^2\Qb_i\;M_{\Qb_f-\Qb_i,n}\;\alpha_{\Qb_i}^i \label{TM}
\end{align}
with
\begin{align}
M_{\Qb_f-\Qb_i}=(-2\pi\ii/\hbar v)\bra{n}\bra{\qb_f}\hat{\mathcal{H}_1}\ket{\qb_i}\ket{0}. \label{MH}
\end{align}
We remark that the transfer-matrix elements defined in Eq.~\eqref{MH} involve just the difference between incident and scattered transverse wave vectors. In what follows, we develop a formalism to relate $M_{\Qb_f-\Qb_i}$ to the EELS probability and obtain specific expressions for plasmonic and atomic-vibration modes.

\section{EELS with shaped electron beams}

We consider the configuration of Fig.~\ref{Fig1}a and assume the electron velocities and sample dimensions to be small enough as to neglect retardation effects and work in the electrostatic regime. Further adopting the aforementioned nonrecoil approximation, we can disregard the longitudinal component of the electron wave function and only consider the dependence on transverse coordinates $\Rb=(x,y)$ (i.e., taking the electron velocity $\vb$ along $z$). We can then write a general expression for the EELS probability $\Gamma_{\rm EELS}(\omega)$ in terms of the energy loss $\hbar\omega$, the transverse wave vector $\Qb_f\perp\zz$ of the final ($f$) electron state (corresponding to a wave function $\ee^{\ii\Qb_f\cdot\Rb}/2\pi$), and the transverse component of the initial ($i$) electron wave function, $\psi_i(\Rb)$. More precisely, using Eq.\ (17) of ref.~\citenum{paper149}, we have $\Gamma_{\rm EELS}(\omega)=\int d^2\Qb_f\;\Gamma_{\rm EELS}(\Qb_f,\omega)$, where
\begin{align}
\Gamma_\mathrm{EELS}(\Qb_f,\omega)=&\frac{e^2}{4\pi^3\hbar v^2}\int d^2\Rb\int d^2\Rb'\;\psi_{i}(\Rb)\psi^*_{i}(\Rb') \nonumber\\
&\times\ee^{\ii\Qb_f\cdot(\Rb'-\Rb)}\mathcal{W}(\Rb,\Rb',\omega) \label{Eq:dGamdQf}
\end{align}
is the momentum-resolved probability and
\begin{align}
\mathcal{W}(\Rb,\Rb',\omega)=\int_{-\infty}^\infty dz&\int_{-\infty}^\infty dz'\,\ee^{\ii\omega(z-z')/v} \label{Eq:W_integrated}\\
&\times\Imm\left\{-W(\rb,\rb',\omega)\right\} \nonumber
\end{align}
is a transverse screened interaction obtained from the full screened interaction $W(\rb,\rb',\omega)$. The latter stands for the Coulomb potential created at $\rb$ by a point charge of magnitude $\ee^{-\ii\omega t}$ placed at $\rb'$, including the effect of screening by the environment. Now, as we show below for plasmonic and phononic structures, the transverse screened interaction in Eq.~\eqref{Eq:W_integrated} is separable as
\begin{align}
\mathcal{W}(\Rb,\Rb',\omega)=\sum_n g_n(\omega) w_n(\Rb,\omega)w_n^\ast(\Rb',\omega),
\label{Eq:W}
\end{align}
where $n$ runs over excitation modes characterized by spatial profiles $w_n(\Rb,\omega)$ and dimensionless spectral functions $g_n(\omega)$. Finally, inserting Eq.~\eqref{Eq:W} into Eq.~\eqref{Eq:dGamdQf}, we readily find Eq.~\eqref{Eq:dGamdQf1} in the main text. Incidentally, the angle-integrated inelastic electron signal (i.e., the integral of Eq.~\eqref{Eq:dGamdQf} over $\Qb_f$) reduces to $\Gamma_\mathrm{EELS}(\omega)=(e^2/\pi\hbar v^2)\sum_n g_n(\omega)\int d^2\Rb\;\big|\psi_{i}(\Rb)\big|^2 \big|w_n(\Rb,\omega)\big|^2$, which is an average over transverse positions $\Rb$ weighted by both the incident electron probability \cite{RH1977,paper149} and the mode spatial profile, and consequently, since the ebeam can generally excite different modes $n$, the optimization scheme that we pursue here to produce entanglement essentially consists in rearranging the $\Qb_f$ distribution of the scattered electron component associated with the excitation of each of those modes.

We note that the spectral functions in this formalism can be generally approximated by Lorentzians,
\begin{align}
g_n(\omega)\approx{\rm Im}\left\{\frac{G_n/\pi}{\omega_n-\omega-\ii\gamma_n/2}\right\},
\nonumber
\end{align}
peaked at the mode energies $\hbar\omega_n$ and having areas $G_n$ and widths $\gamma_n$ (see below) that determine the spectral positions and strengths of the EELS features.

\section{Numerical determination of $\ket{\psi_i^{\rm el}}$ for creating selected excitations and entangled electron-sample states}

Given a desired final state defined through the coefficients $\alpha_{\Qb_f n}^f$, we numerically obtain $\alpha_{\Qb_i}^i$ by inverting Eq.~\eqref{aMa} upon discretization of $\Qb_i$ using a finite number of points (pixels at the electron analyzer in the Fourier plane $\Qb_f$), as specified in the main text. More precisely, we follow a simple procedure consisting in specifying target values of $\alpha_{\Qb_f n}^f$ within a region $Q<Q_{f,{\rm max}}$ (effectively setting it to 0 outside it) and obtain $\alpha_{\Qb_i}^i$ for $Q_i<Q_{i,{\rm max}}$ through the noted numerical inversion. The wave vector ranges are related to the maximum incidence|collection half-angle $\varphi_{i|f}$ through $Q_{i|f,{\rm max}}=(\me v/\hbar)\sin\varphi_{i|f}$. In this procedure, to select a single sample excitation $n=n_0$ (Fig.~\ref{Fig2}), we set $\alpha_{\Qb_f n}^f=C\delta_{nn_0}\Theta(Q_{f,{\rm max}}-Q_f)$, where $C$ is a constant and $\Theta$ is the step function. However, to produce electron-sample entanglement involving two (Fig.~\ref{Fig3}) or three (Fig.~\ref{Fig4}) sample states $n_j$ correlated with final electron wave vectors $\Qb_j$ (see Fig.~\ref{Fig1}b), we set $\alpha_{\Qb_f n_j}^f$ to a constant at the $\Qb_f$-space pixel that contains $\Qb_j$, and zero elsewhere. We then construct $\ket{\psi_i^{\rm el}}$ from the obtained coefficients $\alpha_{\Qb_i}^i$ (also setting them to zero for $Q_i>Q_{i,{\rm max}}$), and insert this input wave function in Eq.~\eqref{Eq:dGamdQf1} to generate the actual final state, plotted in the figures with a finer discretization in $\Qb_f$ space.

\section{Transfer matrix from the spectral and spatial mode functions}

An expression for the EELS probability analogous to Eq.~\eqref{Eq:dGamdQf1} can be readily obtained from Eq.~\eqref{TM}:
\begin{align}
\Gamma_\mathrm{EELS}(\Qb_f,\omega)=\sum_n\left|\int \!d^2\Qb_i\;M_{\Qb_f-\Qb_i,n}\;\alpha_{\Qb_i}^i\right|^2\!\delta(\omega-\omega_n). \label{EELSbis}
\end{align}
The connection between Eqs.~\eqref{Eq:dGamdQf1} and \eqref{EELSbis} is established by adding finite mode widths $\gamma_n$ to the latter and expanding the incident electron wave function in the former as an integral over momentum components, as indicated in Eq.~\eqref{psiiel}. Comparing the two resulting expressions, we find
\begin{align}
M_{\Qb n}=\frac{e}{4\pi^2v}\sqrt{\frac{G_n}{\pi\hbar}}
\int d^2\Rb\;\ee^{\ii\Qb\cdot \Rb}w_n(\Rb,\omega), \label{Mfinal}
\end{align}
which provides a prescription to obtain the transfer-matrix coefficients defined in Eq.~\eqref{MH} directly from the screened interaction, thus bypassing the need for a detailed specification of the interaction Hamiltonian. Then, the spatial profiles in Eq.~\eqref{Mvsw} are simply given by the inverse Fourier transform of Eq.~\eqref{Mfinal}.

\section{Transfer matrix and transverse screened interaction for plasmonic nanoparticles}

In the electrostatic limit under consideration, we can recast the response of an arbitrarily shaped homogeneous nanoparticle into an eigenvalue problem \cite{paper010,BK12}. We then need to find the real eigenvalues $\lambda_n$ and eigenvectors $\sigma_n(\mathbf{s})$ of the integral equation $2\pi\lambda_n\sigma_n(\mathbf{s})=\oint d\mathbf{s}' F(\mathbf{s},\mathbf{s}')\sigma_n(\mathbf{s}'),$ where $\mathbf{s}$ and $\mathbf{s}'$ run over particle surface coordinates, and $F(\mathbf{s},\mathbf{s}')=-\mathbf{n}\cdot(\mathbf{s}-\mathbf{s}')/\lvert \mathbf{s}-\mathbf{s}'\rvert^3$. Here, we solve this eigensystem for triangular particles using the MNPBEM toolbox \cite{HT12}, based on a finite boundary-element discretization of the particle surface.  Then, the spectral functions in Eq.~\eqref{Eq:W} reduce to \cite{paper010,BK12}
\begin{align}
g_n(\omega)={\rm Im}\left\{\frac{-2}{\epsilon (1+\lambda_n)+(1-\lambda_n)}\right\},
\nonumber
\end{align}
whereas the spatial profiles become
\begin{align}
w_n(\Rb,\omega)=2\oint d\mathbf{s}\,\sigma_n(\mathbf{s})\ee^{-\ii \omega s_z/v}K_0\left(\frac{\omega\left\lvert \Rb-\mathbf{S}\right\rvert}{v}\right)
\nonumber
\end{align}
with $\mathbf{s}=\mathbf{S}+s_z\zz$. This expression neglects the contribution of bulk modes, which should be a reasonable approximation at loss energies well below the bulk plasmon. Inserting it into Eq.~\eqref{Mfinal}, the transfer-matrix elements reduce to
\begin{align}
M_{\Qb,n}\approx\frac{e}{2\pi v}\sqrt{\frac{G_n}{\pi\hbar}}\frac{\ee^{\ii\Qb\cdot\Sb}}{Q^2+\omega_n^2/v^2}\oint d\mathbf{s}\,\sigma_n(\mathbf{s})\ee^{-\ii \omega s_z/v},
\nonumber
\end{align}
where we have approximated $\omega\approx\omega_n$. For silver, we model the dielectric function as \cite{paper149} $\epsilon=\epsilon_b-\omega_p^2/\omega(\omega+\ii\gamma)$ with $\epsilonb=4.0$, $\hbar\omega_p=9.17\,$eV, and $\hbar\gamma=21\,$meV, yielding mode frequencies $\omega_n=\omega_p/\sqrt{\epsilonb+(1-\lambda_n)/(1+\lambda_n)}$, flat widths $\gamma_n\approx\gamma$, and spectral weights $G_n=\pi \omega_n^3/[\omega_p^2(1+\lambda_n)]$.

\section{Transfer matrix and transverse screened interaction for atomic vibrations}

For molecules or nanoparticles whose mid-infrared response is dominated by atomic vibrations, we find the spectral and spatial dependence of the modes in Eq.~\eqref{Eq:W} to be governed by \cite{paper263,paper376}
\begin{align}
g_n(\omega)=\Imm\left\lbrace\frac{\omega_n^2}{\omega_n^2-\omega(\omega+\ii\gamma)}\right\rbrace
\label{gnvib}
\end{align}
and
\begin{align} % the units of w_n must be L^0.5
w_n(\Rb,\omega)=\frac{2}{\omega_n}\sum_l\frac{1}{\sqrt{M_l}}&\int\,d^3\rb'\;K_0(\omega|\Rb-\Rb'|/v) \nonumber\\
&\times\ee^{\ii \omega z'/v}\,\left[\eb_{nl}\cdot\vec{\rho}_l(\rb')\right],
\label{Eq:wn_vib}
\end{align}
where $n$ now runs over vibrational modes, $\omega_n$ and $\eb_{nl}$ are the corresponding real frequencies and normalized atomic displacement vectors ($\sum_l\eb_{nl}\cdot\eb_{n'l}=\delta_{nn'}$), respectively, the $l$ sum extends over the atoms in the structure, $M_l$ is the mass of atom $l$, $\vec{\rho}_l(\rb)$ denotes the gradient of the charge distribution associated with displacements of that atom, and we have incorporated a phenomenological damping rate $\gamma$ (here set to $\hbar\gamma=1\,$meV). From Eq.~\eqref{gnvib}, we have $\gamma_n\approx\gamma$ for all modes, as well as $G_n\approx\pi\omega_n/2$. Following ref.~\citenum{paper376}, we use density-functional theory (DFT) to calculate $\vec{\rho}_l(\rb)$, $\omega_n$, and $\eb_{nl}$ (see below). The prescription $|\Rb-\Rb'|\rightarrow\sqrt{|\Rb-\Rb'|^2+\Delta^2}$ is also adopted with $\Delta=0.2\,${\AA} to approximately account for a cutoff $\sim\hbar/\Delta$ in momentum transfer \cite{paper149} and so avoid the unphysical divergence associated with close electron-atom encounters.

\section{First-principles description of atomic vibrations}

We use DFT and the projector-augmented-wave method \cite{B94_2} as implemented in the Vienna \textit{ab initio} simulation package \cite{KF96,KH93,KF96_2} (VASP) with the Perdew-Burke-Ernzerhof generalized gradient approximation for electron exchange and correlation \cite{PBE96}. We apply this method to describe hBN flakes with hydrogen-passivated edges, using a plane-wave cutoff energy of 500\,eV, as well as a sufficient amount of vacuum spacing in all directions around the structure to avoid interaction among the periodic images. Atomic equilibrium positions are found by minimizing the total energy using the conjugate gradient method with convergence criteria between consecutive iteration steps set to $10^{-5}$\,eV for the total energy and 0.02\,eV/{\AA} for the atomic forces. Vibrational frequencies and eigenmodes are found by diagonalizing the dynamical matrix, which is calculated for 0.01\,{\AA} displacements. The corresponding gradients $\vec{\rho}_l^{\,\,\rm val}(\rb)$ of the charge distribution are obtained by treating core electrons and nuclei as point particles, while the contribution coming from valence electrons is directly taken from DFT using a dense grid.

%\section*{References}

\section*{Acknowledgments}
This work has been supported in part by the European Research Council (Advanced Grant 789104-eNANO), the Spanish MICINN (PID2020-112625GB-I00 and Severo Ochoa CEX2019-000910-S), the Catalan CERCA Program, and Fundaci\'{o}s Cellex and Mir-Puig. AK was supported by the ESF under the project CZ.02.2.69/0.0/0.0/20-079/0017436.

\end{document}